# Multigrid Iterative Algorithm based on Compact Finite Difference Schemes and Hermite interpolation for Solving Regime Switching American Options

**Chinonso I. Nwankwo[a,*], Weizhong Dai[b]**

[a] Department of *Mathematics,* Statistics, and Computer Science, University of Illinois at Chicago, Chicago, IL 60607, USA

[b] Department of Mathematics and Statistics, Louisiana Tech University, Ruston LA 71272, USA
[*] Corresponding author, nonsonwankwo@gmail.com

## Abstract

We present a multigrid iterative algorithm for solving a system of coupled free boundary problems for pricing American put options with regime-switching. The algorithm is based on our recently developed compact finite difference scheme coupled with Hermite interpolation for solving the coupled partial differential equations consisting of the asset option and the delta, gamma, and speed sensitivities. In the algorithm, we first use the Gauss-Seidel method as a smoother and then implement a multigrid strategy based on modified cycle (M-cycle) for solving our discretized equations. Hermite interpolation with Newton interpolatory divided difference (as the basis) is used in estimating the coupled asset, delta, gamma, and speed options in the set of equations. A numerical experiment is performed with the two- and four- regime examples and compared with other existing methods to validate the optimal strategy. Results show that this algorithm provides a fast and efficient tool for pricing American put options with regime-switching.

## 1. Introduction

American style option valuation based on the regime-switching model involves a coupled free boundary problem. The closed-form of the solution for this model is difficult to find, hence, several numerical methods have been proposed [32,23,50,7,18,12,19,27,41,42,25,39,47]. Moreover, it is always a challenging task to approximate the Greeks associated with this model especially, beyond two-regime.

In our previous work [33], we generated a system of coupled partial differential equations through logarithmic transformation and by taking further derivatives of the regime-switching model. We called this system of equations, the asset-delta-gamma-speed equations. By employing the compact finite difference method for discretizing the system of coupled equation and the Hermite interpolation for estimating the coupled regime, we were able to approximate the asset option, option Greeks, and optimal exercise boundary for each regime with high order accuracy using Gauss-Seidel iterative methods.

Although the numerical scheme provides an accurate solution, we have encountered a computational burden when using the Gauss-Seidel iteration. This effect is much more substantial when small grid size and time step are used. Thus, there is a need to find a strategy that reduces the number of iterations and CPU time required to achieve numerical convergence. One possible way is to use the method in Egorova et al. [12]. The authors treated the coupled regime explicitly and the non-coupled part implicitly to reduce the computational cost when using an implicit finite difference scheme. Unfortunately, we found that such implementation reduces the accuracy of our numerical solutions, particularly, for the case beyond the two-regime example. Another possible way is to develop a multigrid iterative algorithm to speed up our computation for solving the system of equations while preserving the accuracy of our numerical solutions.

As we know, the major advantage of the multigrid iterative method is that it can remove the high-frequency error from numerical solutions in a few iterations (using a better smoother) and transfers the low-frequency error to a coarse grid where the latter is cheaper and faster to remove. This improves the overall computational effort, speed, and efficiency of numerical approximations for large-scale problems. Because of this, the multigrid method has gained a broader application after the work of Fedorenko [14], Bachvalov [3], Brandt [4], and Saad [37] which has been applied to elliptic and time-dependent partial differential equations [45,36,20,26,46,24,34,49,31,15,40,16]. For option pricing, Urschel [44] implemented an adaptive space-time multigrid approach to barrier options using the implicit Euler and Crank-Nicholson method. Jeong et al. [21] applied an adaptive multigrid method on the Black Scholes equation with Crank-Nicholson discretization. Clarke and Parrott [9] implemented a multigrid method in discrete linear complementary problem for pricing American options with stochastic volatility.

In this study, we will implement the multigrid method to a compact finite difference scheme coupled with the Hermite interpolation for pricing American options with regime-switching. The rest of the paper is organized as follows. In section 2, we present our mathematical model. In section 3, we present the compact finite difference scheme and estimate the coupled asset, delta, gamma, and speed options in the set of equations for each regime using the Hermite interpolation with the Newton interpolatory divided difference formula as the basis. In section 4, we implement the multigrid strategy and its algorithm for solving the derived numerical scheme and obtain the option values, optimal exercise boundary, and the Greeks in each regime. In section 5, we investigate and compare the numerical performance of our present algorithm using the two- and four-regime example and conclude the paper in section 6.

## 2. Mathematical Model

Our mathematical model is based on the American put option written on the asset $S_t$ with strike price $K$ and expiration time $T$. Let $V_m(S,t)$ denote the option price in the $m^{th}$ regime and $\tau = T - t$ where $m = 1,2,\cdots,I$, and $I$ is the number of regimes. Then, $V_m(S,\tau)$ satisfies the coupled free boundary value problem:

$$-\frac{\partial V_m(S,\tau)}{\partial \tau} + \frac{1}{2}\sigma^2{}_m S^2 \frac{\partial^2 V_m(S,\tau)}{\partial S^2} + r_m S \frac{\partial V_m(S,\tau)}{\partial S} - r_m V_m(S,\tau) + \sum_{l \neq m} q_{ml}\,[V_l(S,\tau) - V_m(S,\tau)] = 0,$$

$$\text{for } S > s_{f(m)}(\tau), \tag{1a}$$

$$V_m(S,\tau) = K - S, \qquad \text{for } S < s_{f(m)}(\tau). \tag{1b}$$

Here, the initial and boundary conditions are given as:

$$V_m(S,0) = max(K - S, 0), \qquad s_{f(m)}(0) = K; \tag{2a}$$

$$V_m\big(s_{f(m)},\tau\big) = K - s_{f(m)}(\tau), \qquad V_m(\infty,\tau) = 0, \quad \frac{\partial}{\partial S}V_m\big(s_{f(m)},\tau\big) = -1, \tag{2b}$$

where $s_{f(m)}(\tau)$ is the optimal exercise boundary for the $m^{th}$ regime, $r_m$and $\sigma_m$ are the interest rates and volatilities in each regime, respectively. Here, $q_{ml}$ is the entry elements of the generator matrix $Q_{I\times I}$ with $m,l = 1,2,\cdots,I$, which satisfies the condition below:

$$q_{mm} = -\sum_{l \neq m} q_{ml}, \quad q_{ml} \geq 0, \quad for\ l \neq m, \quad l = 1,2,\cdots,I. \tag{3}$$

We further fix the free boundary challenge by employing a front-fixing logarithmic transformation [38,12] on multi-variable domains as

$$x_m = \ln\frac{S}{s_{f(m)}(\tau)} = \ln S - \ln s_{f(m)}(\tau), \qquad m = 1,2,\cdots,I. \tag{4a}$$

The transformed option value function $U_m(x_m,\tau)$ is related to the original option value function $V_m(S,t)$ by the transformation

$$U_m(x_m,\tau) = V_m(S,\tau), \qquad m = 1,2,\cdots,I. \tag{4b}$$

Using this transformation and eliminating the first-order derivative by taking further derivatives, we then obtain a set of coupled partial differential equations consisting of the asset, delta, gamma, and speed options as follows:

$$\frac{\partial U_m}{\partial \tau} - \frac{1}{2}{\sigma^2}_m \frac{\partial^2 U_m}{\partial x_m^2} - \left(\frac{{s'}_{f(m)}}{s_{f(m)}} + r_m - \frac{{\sigma^2}_m}{2}\right) W_m + (r_m - q_{mm}) U_m - \sum_{l \neq m} q_{ml} U_l = 0, \tag{5a}$$

$$\frac{\partial W_m}{\partial \tau} - \frac{1}{2}{\sigma^2}_m \frac{\partial^2 W_m}{\partial x_m^2} - \left(\frac{{s'}_{f(m)}}{s_{f(m)}} + r_m - \frac{{\sigma^2}_m}{2}\right) \frac{\partial^2 U_m}{\partial x_m^2} + (r_m - q_{mm}) W_m - \sum_{l \neq m} q_{ml} W_l = 0, \tag{5b}$$

$$\frac{\partial Y_m}{\partial \tau} - \frac{1}{2}{\sigma^2}_m \frac{\partial^2 Y_m}{\partial x_m^2} - \left(\frac{{s'}_{f(m)}}{s_{f(m)}} + r_m - \frac{{\sigma^2}_m}{2}\right) \frac{\partial^2 W_m}{\partial x_m^2} + (r_m - q_{mm}) Y_m - \sum_{l \neq m} q_{ml} Y_l = 0, \tag{5c}$$

$$\frac{\partial Z_m}{\partial \tau} - \frac{1}{2}{\sigma^2}_m \frac{\partial^2 Z_m}{\partial x_m^2} - \left(\frac{{s'}_{f(m)}}{s_{f(m)}} + r_m - \frac{{\sigma^2}_m}{2}\right) \frac{\partial^2 Y_m}{\partial x_m^2} + (r_m - q_{mm}) Z_m - \sum_{l \neq m} q_{ml} Z_l = 0, \tag{5d}$$

where $m, l = 1, 2, \cdots, I$, $x_m \in [0, \infty)$, and $W_m, Y_m$, and $Z_m$ are delta, gamma, and speed options, respectively. The initial and boundary conditions for $U_m(x_m, \tau)$, $W_m(x_m, \tau)$, $Y_m(x_m, \tau)$, and $Z_m(x_m, \tau)$ are given as:

$$s_{f(m)}(0) = K, \qquad U_m(x_m, 0) = 0, \qquad W_m(x_m, 0) = 0, \qquad Y_m(x_m, 0) = 0, \qquad Z_m(x_m, 0) = 0; \tag{6a}$$

$$U_m(0, \tau) = K - s_{f(m)}(\tau), \qquad W_m(0, \tau) = -s_{f(m)}(\tau); \tag{6b}$$

$$Y_m(0, \tau) = -s_{f(m)}(\tau), \qquad Z_m(0, \tau) = -s_{f(m)}(\tau); \tag{6c}$$

$$U_m(\infty, \tau) = 0; \;\; W_m(\infty, \tau) = 0; \;\; Y_m(\infty, \tau) = 0, \quad Z_m(\infty, \tau) = 0. \tag{6d}$$

**Remark 1:** A similar system of PDEs as in (5a)-(5d) has been used in the work of Liao and Khaliq [28] and Dremkova and Ehrhardt [10] for solving the European option and obtaining only the asset and delta options. Here, we use the set of four equations in (5) that have the advantages as follows:

- To approximate the Greeks simultaneously with the asset option and optimal exercise boundary using the fourth-order compact operator. Approximating the Greeks from the numerical solutions of the asset option could further introduce errors due to the sensitivity of the Greek parameters.
- The first derivative that introduces nonlinearity in the model due to the presence of the optimal exercise boundary and its derivative is removed as shown in (5a)-(5d). This enables us to treat this nonlinear coefficient in the force term after discretization and present a linear system in (17) and (19).

**Remark 2**: $Z_m$ represents the speed option for each regime. The speed option is the rate of change of gamma with respect to the stock price. It is one of the useful Greeks to monitor when delta-hedging or gamma hedging a portfolio.

## 3. Numerical Discretization and Interpolation.

To solve the above system of PDEs that consist of the asset, delta, gamma, and speed options in a uniform grid $[0, \infty) \times [0\ T]$ for each regime and take into account the relationship among the regimes' intervals using cubic interpolation, the infinite domain is truncated to a finite large domain with an estimated boundary $(x_m)_M$ far enough that the error generated due to the truncation is negligible [12,22]. Letting $i$ and $j$ represent the node points in the $l^{th}$ and $m^{th}$ regimes' intervals, respectively, and M and N represent the numbers of grid points and time steps, respectively, we define

$$(x_m)_i = ih, \qquad (x_l)_j = jh, \qquad h = \frac{(x_m)_M}{M} = \frac{(x_l)_M}{M}, \qquad i, j\ \in [0, \mathrm{M}]; \tag{7a}$$

$$k = \frac{T}{N}, \qquad \tau_n = nk, \qquad k \in [0, \mathrm{N}]. \tag{7b}$$

Here, we choose the same length of interval for all regimes so that we have a uniform grid size. In this section, we find the numerical solutions of the asset option and optimal exercise boundary for each regime which we denote as $(u_m)_i^n$, and $s_{f(m)}^n$, respectively. We will further obtain the numerical solutions of the option Greeks $(w_m)_i^n, \left(y_m\right)_i^n,\ (z_m)_i^n, \Theta_m, \mathrm{K}_m,$ and $\Gamma_m$ simultaneously with the asset option and optimal exercise boundary in each regime, where $(w_m)_i^n, \left(y_m\right)_i^n,\ (z_m)_i^n, \Theta_m, \mathrm{K}_m,$ and $\Gamma_m$ denote the delta, gamma, speed, theta, delta decay, and color options, respectively.

### 3.1. Compact Finite Difference Scheme

Based on our previous study, the set of equations in (5) will be discretized using compact and Crank-Nicolson scheme in space and time, respectively. To develop a compact finite difference scheme in space for the asset option at $(x_m)_0 = 0$, we first employ a compact finite difference formula as described in the following lemma.

**Lemma.** Assume $f(x) \in C^6[x_0, x_1]$, then it holds

$$\frac{7}{4}f''(x_0) + \frac{3}{4}f''(x_1) = \frac{5}{h^2}[f(x_1) - f(x_0)] - \frac{5}{h}f'(x_0) - \frac{h}{4}f^{(3)}(x_0) + \frac{h}{6}f^{(3)}(x_1) + O(h^4). \tag{8a}$$

**Proof.** It is straightforward by using the Taylor series expansion.

Furthermore, to shift the third derivative of the asset option in (8a) away from the boundary point, we implement a fourth-order compact approximation as follows:

$$f^{(3)}(x_0) = \frac{12}{h}[f'(x_0) - 2f'(x_1) + f'(x_2)] - 10f^{(3)}(x_1) - f^{(3)}(x_2) + O(h^4). \qquad (8b)$$

Substituting $f^{(3)}(x_0)$ in (8b) into (8a), we obtain as follows:

$$\frac{7}{4}f''(x_0) + \frac{3}{4}f''(x_1) = \frac{5}{h^2}[f(x_1) - f(x_0)] - \frac{1}{h}[8f'(x_0) - 6f'(x_1) + 3f'(x_2)] + \frac{32h}{12}f^{(3)}(x_1) + \frac{3h}{12}f^{(3)}(x_2) + O(h^4). \qquad (8c)$$

Using (8c) for the second-order derivative term in (1a), we obtained

$$\frac{1}{2}\sigma^2{}_m\left[\frac{7}{4}\frac{\partial^2 U_m\big((x_m)_0,\tau_{n+1/2}\big)}{\partial x_m^2} + \frac{3}{4}\frac{\partial^2 U_m\big((x_m)_1,\tau_{n+1/2}\big)}{\partial x_m^2}\right]$$
$$= \frac{5\sigma^2{}_m}{2}\left[\frac{U_m\big((x_m)_1,\tau_{n+1/2}\big) - U_m\big((x_m)_0,\tau_{n+1/2}\big)}{h^2} - \frac{1}{h}\frac{\partial U_m\big((x_m)_0,\tau_{n+1/2}\big)}{\partial x_m}\right]$$
$$- \frac{3\sigma^2{}_m}{2h}\left[\frac{\partial U_m\big((x_m)_0,\tau_{n+1/2}\big)}{\partial x_m} - 2\frac{\partial U_m\big((x_m)_1,\tau_{n+1/2}\big)}{\partial x_m} + \frac{\partial U_m\big((x_m)_2,\tau_{n+1/2}\big)}{\partial x_m}\right]$$
$$- \frac{\sigma^2{}_m h}{2}\left[\frac{32}{12}\frac{\partial^3 U_m\big((x_m)_1,\tau_{n+1/2}\big)}{\partial x_m^3} + \frac{3}{12}\frac{\partial^3 U_m\big((x_m)_2,\tau_{n+1/2}\big)}{\partial x_m^3}\right] + O(h^4). \qquad (9)$$

The first-order derivative in (9) at $(x_m)_0$ is then evaluated based on (6b) as

$$\frac{\partial U_m\big((x_m)_0,\tau_{n+1/2}\big)}{\partial x_m} - U_m\big((x_m)_0,\tau_{n+1/2}\big) = W_m\big((x_m)_0,\tau_{n+1/2}\big) - U_m\big((x_m)_0,\tau_{n+1/2}\big) = -K. \qquad (10)$$

Next, we evaluate the third-order derivative term in (9) and discretize $\partial W_m/\partial t$. This gives

$$\frac{32\sigma^2{}_m h}{24}\frac{\partial^3 U_m\big((x_m)_0,\tau_{n+1/2}\big)}{\partial x_m^3} + \frac{3\sigma^2{}_m h}{24}\frac{\partial^3 U_m\big((x_m)_2,\tau_{n+1/2}\big)}{\partial x_m^3}$$
$$= \frac{32h}{12}\left[\frac{W_m((x_m)_1,\tau_{n+1}) - W_m((x_m)_1,\tau_n)}{k}\right] + \frac{3h}{12}\left[\frac{W_m((x_m)_2,\tau_{n+1}) - W_m((x_m)_2,\tau_n)}{k}\right]$$
$$- \frac{32h}{12}(\beta_m)_{n+1/2}Y_m\big((x_m)_1,\tau_{n+1/2}\big) - \frac{3h}{12}(\beta_m)_{n+1/2}Y_m\big((x_m)_2,\tau_{n+1/2}\big)$$
$$+ \frac{32h}{12}(r_m - q_{mm})W_m\big((x_m)_1,\tau_{n+1/2}\big) + \frac{3h}{12}(r_m - q_{mm})W_m\big((x_m)_2,\tau_{n+1/2}\big)$$
$$- \frac{32h}{12}\sum_{l\neq m} q_{ml}\,W_l\big((x_m)_{j^*|i=1},\tau_{n+1/2}\big) - \frac{3h}{12}\sum_{l\neq m} q_{ml}\,W_l\big((x_m)_{j^*|i=2},\tau_{n+1/2}\big) + O(k^2). \qquad (11)$$

where

$$(\beta_m)_{n+1/2} \equiv \frac{2\left(s^{n+1}_{f(m)} - s^{n}_{f(m)}\right)}{k\left(s^{n+1}_{f(m)} + s^{n}_{f(m)}\right)} + r_m - \frac{\sigma^2{}_m}{2}. \quad (12)$$

Here, $(x_m)_{j^*|i=1}$ and $(x_m)_{j^*|i=2}$ are the locations in the space for the $l^{th}$ equation corresponding to $(x_m)_1$ and $(x_m)_2$ in the $m^{th}$ equation, respectively.

Substituting (11) and (13) into (9) and applying the Crank-Nicholson method in time, we then obtain the compact finite difference scheme at $x_m = 0$ for the asset option as

$$\begin{aligned}
&\frac{7}{4}\left[\frac{(u_m)_0^{n+1} - (u_m)_0^n}{k}\right] + \frac{3}{4}\left[\frac{(u_m)_1^{n+1} - (u_m)_1^n}{k}\right] - \frac{5\sigma^2{}_m}{4h}\left[\frac{(u_m)_1^{n+1} - (u_m)_0^{n+1}}{h} - (u_m)_0^{n+1}\right] \\
&\quad - \frac{5\sigma^2{}_m}{4h}\left[\frac{(u_m)_1^{n} - (u_m)_0^{n}}{h} - (u_m)_0^{n}\right] - \frac{5\sigma^2{}_m}{2h}K \\
&\quad + \frac{3\sigma^2{}_m}{4}\left[\frac{[(w_m)_0^{n+1} + (w_m)_0^n] - 2[(w_m)_1^{n+1} + (w_m)_1^n] + [(w_m)_2^{n+1} + (w_m)_2^n]}{h}\right] \\
&\quad + \frac{(r_m - q_{mm})}{8}\left[7[(u_m)_0^{n+1} + (u_m)_0^n] + 3[(u_m)_1^{n+1} + (u_m)_1^n]\right] \\
&\quad - \frac{h}{12}\left[\frac{32[(w_m)_1^{n+1} - (w_m)_1^n] + 3[(w_m)_2^{n+1} - (w_m)_2^n]}{k}\right] \\
&\quad - \frac{h(r_m - q_{mm})}{24}\left[32[(w_m)_1^{n+1} + (w_m)_1^n] + 3[(w_m)_2^{n+1} + (w_m)_2^n]\right] \\
&\quad - \frac{(\beta_m)_{n+1/2}}{8}\left[7[(w_m)_0^{n+1} + (w_m)_0^n] + 3[(w_m)_1^{n+1} + (w_m)_1^n]\right] \\
&\quad + \frac{h(\beta_m)_{n+1/2}}{24}\left[32[(y_m)_1^{n+1} + (y_m)_1^n] + 3[(y_m)_2^{n+1} + (y_m)_2^n]\right] \\
&\quad + \frac{h}{24}\sum_{l\neq m} q_{ml}\left[32(w_l)_{j^*|i=1}^{n+1} + (w_l)_{j^*|i=1}^{n} + 3\left((w_l)_{j^*|i=2}^{n+1} + (w_l)_{j^*|i=2}^{n}\right)\right] \\
&\quad - \frac{1}{8}\sum_{l\neq m} q_{ml}\left[7\left((u_l)_{j^*|i=0}^{n+1} + (u_l)_{j^*|i=0}^{n}\right) + 3\left((u_l)_{j^*|i=1}^{n+1} + (u_l)_{j^*|i=1}^{n}\right)\right] = 0, \quad (13)
\end{aligned}$$

which is further simplified to

$$a_m^1 (u_m)_0^{n+1} + b_m^1 (u_m)_1^{n+1} = (f_0^u)_m^{n+1/2}\,, \quad (14)$$

with the truncation error of $O(k^2 + h^4)$. Here, $j^*|i = 0,1,2$ indicate the values of $j^*$ given at $(x_m)_0$, $(x_m)_1$, and $(x_m)_2$, respectively.

**Remark 3**: It is important to mention that for $\tau > 0$, the asset option $U_m$ is $C^1$ smooth and the second derivative is discontinuous at $x_m = 0$ which contradicts our assumption in (8). However, we slightly shift away from the boundary (at $x_m = 0$) with fourth-order compact approximation when computing the third derivative of the asset option. Moreover, the third derivative of the asset option at $x_m = 0$ which is the highest order derivative involved in our discretize boundary equation is evaluated using (5b).

At each interior grid point, $(x_m)_i = 1,2,\dots,M-1$, using the compact finite difference scheme (Yan et al., 2019) as

$$\frac{1}{12}f''(x_{i-1}) + \frac{10}{12}f''(x_i) + \frac{1}{12}f''(x_{i+1}) = \frac{1}{h^2}[f(x_{i-1}) - 2f(x_i) + f(x_{i+1})] + O(h^4), \tag{15}$$

for (5a)-(5d), we obtain similar finite difference equations as

$$d_m^1(u_m)_{i-1}^{n+1} + c_m^1(u_m)_i^{n+1} + d_m^1(u_m)_{i+1}^{n+1} = (f_i^u)_m^{n+1/2}, \tag{16a}$$

$$d_m^1(w_m)_{i-1}^{n+1} + c_m^1(w_m)_i^{n+1} + d_m^1(w_m)_{i+1}^{n+1} = (f_i^w)_m^{n+1/2}, \tag{16b}$$

$$d_m^1(y_m)_{i-1}^{n+1} + c_m^1(y_m)_i^{n+1} + d_m^1(y_m)_{i+1}^{n+1} = \left(f_i^y\right)_m^{n+1/2}, \tag{16c}$$

$$d_m^1(z_m)_{i-1}^{n+1} + c_m^1(z_m)_i^{n+1} + d_m^1(z_m)_{i+1}^{n+1} = (f_i^z)_m^{n+1/2}, \qquad i = 1,2,\dots,N-1, \tag{16d}$$

where $c_m^1$ and $d_m^1$ are coefficients given later in (20). To further ensure that the entries of our matrix preserve both symmetricity and positive definiteness, we could further multiply (14) by $d_m^1/\ b_m^1$ to obtain

$$\bar{a}_m^1(u_m)_0^{n+1} + d_m^1(u_m)_1^{n+1} = \left(\bar{f}_0^u\right)_m^{n+1/2}, \tag{16e}$$

where

$$\bar{a}_m^1 = \frac{a_m^1 d_m^1}{b_m^1}, \qquad \bar{f}_0^u = \frac{d_m^1 (f_0^u)_m^{n+1/2}}{b_m^1}. \tag{16f}$$

Note that these discrete option Greeks at the right end boundary are zero. Hence, we obtain a discretized linear system for the asset, delta, gamma, and speed option equations as follows:

$$A_m\boldsymbol{u}_m = \boldsymbol{f}_m^u, \qquad B_m\boldsymbol{w}_m = \boldsymbol{f}_m^w, \qquad B_m\boldsymbol{y}_m = \boldsymbol{f}_m^y, \qquad B_m\boldsymbol{z}_m = \boldsymbol{f}_m^z, \tag{17}$$

where

$$A_m = \begin{bmatrix} \bar{a}_m^1 & d_m^1 & & & & & & \\ d_m^1 & c_m^1 & d_m^1 & & & & & \\ & d_m^1 & c_m^1 & d_m^1 & & & & \\ & & d_m^1 & c_m^1 & d_m^1 & & & \\ & & & \ddots & \ddots & \ddots & & \\ & & & & d_m^1 & c_m^1 & d_m^1 & \\ & & & & & d_m^1 & c_m^1 & d_m^1 \\ & & & & & & d_m^1 & c_m^1 \end{bmatrix} \qquad B_m = \begin{bmatrix} c_m^1 & d_m^1 & & & & & & \\ d_m^1 & c_m^1 & d_m^1 & & & & & \\ & d_m^1 & c_m^1 & d_m^1 & & & & \\ & & d_m^1 & c_m^1 & d_m^1 & & & \\ & & & \ddots & \ddots & \ddots & & \\ & & & & d_m^1 & c_m^1 & d_m^1 & \\ & & & & & d_m^1 & c_m^1 & d_m^1 \\ & & & & & & d_m^1 & c_m^1 \end{bmatrix}$$

$$
\boldsymbol{f}_m^u = \begin{bmatrix} (\bar{f}_0^u)_m^{n+1/2} \\ (f_1^u)_m^{n+1/2} \\ (f_2^u)_m^{n+1/2} \\ \cdot \\ \cdot \\ \cdot \\ (f_{M-1}^u)_m^{n+1/2} \end{bmatrix} \quad \boldsymbol{f}_m^w = \begin{bmatrix} (f_1^w)_m^{n+1/2} - b_m^1 (w_m)_0^{n+1} \\ (f_2^w)_m^{n+1/2} \\ (f_3^w)_m^{n+1/2} \\ \cdot \\ \cdot \\ \cdot \\ (f_{M-1}^w)_m^{n+1/2} \end{bmatrix} \quad \cdots \quad \boldsymbol{f}_m^z = \begin{bmatrix} (f_1^z)_m^{n+1/2} - b_m^1 (z_m)_0^{n+1} \\ (f_2^z)_m^{n+1/2} \\ (f_3^z)_m^{n+1/2} \\ \cdot \\ \cdot \\ \cdot \\ (f_{M-1}^z)_m^{n+1/2} \end{bmatrix}. \quad (18)
$$

Here,

$$
\begin{aligned}
(f_0^u)_m^{n+1/2} = {} & a_m^2 (u_m)_0^n + b_m^2 (u_m)_1^n + \frac{5\sigma^2{}_m}{2h} K \\
& - \frac{3\sigma^2{}_m}{4} \left[ \frac{\left[(w_m)_0^{n+1} + (w_m)_0^n\right] - 2\left[(w_m)_1^{n+1} + (w_m)_1^n\right] + \left[(w_m)_2^{n+1} + (w_m)_2^n\right]}{h} \right] \\
& + \frac{h}{12} \left[ \frac{32\left[(w_m)_1^{n+1} - (w_m)_1^n\right] + 3\left[(w_m)_2^{n+1} - (w_m)_2^n\right]}{k} \right] \\
& + \frac{h\left(r_m - q_{mm}\right)}{24} \left[ 32\left[(w_m)_1^{n+1} + (w_m)_1^n\right] + 3\left[(w_m)_2^{n+1} + (w_m)_2^n\right] \right] \\
& + \frac{(\beta_m)_{n+1/2}}{8} \left[ 7\left[(w_m)_0^{n+1} + (w_m)_0^n\right] + 3\left[(w_m)_1^{n+1} + (w_m)_1^n\right] \right] \\
& - \frac{h(\beta_m)_{n+1/2}}{24} \left[ 32\left[\left(y_m\right)_1^{n+1} + \left(y_m\right)_1^n\right] + 3\left[\left(y_m\right)_2^{n+1} + \left(y_m\right)_2^n\right] \right] \\
& - \frac{h}{24} \sum_{l \neq m} q_{ml} \left[ 32 (w_l)_{j^*|i=1}^{n+1} + (w_l)_{j^*|i=1}^n + 3\left( (w_l)_{j^*|i=2}^{n+1} + (w_l)_{j^*|i=2}^n \right) \right] \\
& + \frac{1}{8} \sum_{l \neq m} q_{ml} \left[ 7\left( (u_l)_{j^*|i=0}^{n+1} + (u_l)_{j^*|i=0}^n \right) + 3\left( (u_l)_{j^*|i=1}^{n+1} + (u_l)_{j^*|i=1}^n \right) \right] = 0, \quad (19a)
\end{aligned}
$$

$$
\begin{aligned}
(f_i^u)_m^{n+1/2} = {} & d_m^2 (u_m)_{i-1}^n + c_m^2 (u_m)_i^n + d_m^2 (u_m)_{i+1}^n \\
& + \frac{\mu_m (\beta_m)_{n+1/2}}{24} \left[ [(w_m)_{i-1}^{n+1} + (w_m)_{i-1}^n] + 10[(w_m)_i^{n+1} + (w_m)_i^n] + [(w_m)_{i+1}^{n+1} + (w_m)_{i+1}^n] \right] \\
& + \frac{k}{24} \sum_{l \neq m} q_{ml} \left[ \left[(u_l)_{j^*-1}^{n+1} + (u_l)_{j^*-1}^n\right] + 10\left[(u_l)_{j^*}^{n+1} + (u_l)_{j^*}^n\right] + \left[(u_l)_{j^*+1}^{n+1} + (u_l)_{j^*+1}^n\right] \right], \quad (19b)
\end{aligned}
$$

$$
\begin{aligned}
(f_i^w)_m^{n+1/2} = {} & d_m^2 (w_m)_{i-1}^n + c_m^2 (w_m)_i^n + d_m^2 (w_m)_{i+1}^n \\
& + \frac{\mu_m}{2} (\beta_m)_{n+1/2} \left[ [(u_m)_{i-1}^{n+1} + (u_m)_{i-1}^n] - 2[(u_m)_i^{n+1} + (u_m)_i^n] + [(u_m)_{i+1}^{n+1} + (u_m)_{i+1}^n] \right] \\
& + \sum_{l \neq m} \frac{q_{ml}}{24} \left[ \left[(w_l)_{j^*-1}^{n+1} + (w_l)_{j^*-1}^n\right] + 10\left[(w_l)_{j^*}^{n+1} + (w_l)_{j^*}^n\right] + \left[(w_l)_{j^*+1}^{n+1} + (w_l)_{j^*+1}^n\right] \right], \quad (19c)
\end{aligned}
$$

$$\left(f_i^y\right)_m^{n+1/2} = d_m^2 (y_m)_{i-1}^n + c_m^2 (y_m)_i^n + d_m^2 (y_m)_{i+1}^n$$
$$+\frac{\mu_m}{2}(\beta_m)_{n+1/2}\left[[(w_m)_{i-1}^{n+1} + (w_m)_{i-1}^n] - 2[(w_m)_i^{n+1} + (w_m)_i^n] + [(w_m)_{i+1}^{n+1} + (w_m)_{i+1}^n]\right]$$
$$+\frac{k}{24}\sum_{l\neq m} q_{ml}\left[\left[(y_l)_{j^*-1}^{n+1} + (y_l)_{j^*-1}^n\right] + 10\left[(y_l)_{j^*}^{n+1} + (y_l)_{j^*}^n\right] + \left[(y_l)_{j^*+1}^{n+1} + (y_l)_{j^*+1}^n\right]\right], \quad (19d)$$

$$(f_i^z)_m^{n+1/2} = d_m^2 (z_m)_{i-1}^n + c_m^2 (z_m)_i^n + d_m^2 (z_m)_{i+1}^n$$
$$+\frac{\mu_m}{2}(\beta_m)_{n+1/2}\left[[(y_m)_{i-1}^{n+1} + (y_m)_{i-1}^n] - 2[(y_m)_i^{n+1} + (y_m)_i^n] + [(y_m)_{i+1}^{n+1} + (y_m)_{i+1}^n]\right]$$
$$+\frac{k}{24}\sum_{l\neq m} q_{ml}\left[\left[(z_l)_{j^*-1}^{n+1} + (z_l)_{j^*-1}^n\right] + 10\left[(z_l)_{j^*}^{n+1} + (z_l)_{j^*}^n\right] + \left[(z_l)_{j^*+1}^{n+1} + (z_l)_{j^*+1}^n\right]\right], \quad (19e)$$

$$\mu_m = \frac{\sigma^2{}_m k}{h^2}, \qquad a_m^1 = \frac{7}{4} + \frac{5}{4}\mu_m + \frac{5h}{4}\mu_m + \frac{7k}{8}(r_m - q_{mm}), \qquad b_m^1 = \frac{3}{4} - \frac{5}{4}\mu_m + \frac{3k}{8}(r_m - q_{mm}); \quad (20a)$$

$$c_m^1 = \frac{10}{12} + \frac{\mu_m}{2} + \frac{10k}{24}(r_m - q_{mm}), \qquad d_m^1 = \frac{1}{12} - \frac{\mu_m}{4} + \frac{k}{24}(r_m - q_{mm}); \quad (20b)$$

$$a_m^2 = \frac{7}{4} - \frac{5}{4}\mu_m - \frac{5h}{4}\mu_m - \frac{7k}{8}(r_m - q_{mm}), \qquad b_m^2 = \frac{3}{4} + \frac{5}{4}\mu_m - \frac{3k}{8}(r_m - q_{mm}); \quad (20c)$$

$$c_m^2 = \frac{10}{12} - \frac{\mu_m}{2} - \frac{10k}{24}(r_m - q_{mm}), \qquad d_m^2 = \frac{1}{12} + \frac{\mu_m}{4} - \frac{k}{24}(r_m - q_{mm})\,; \quad (20d)$$

where $j^*$ represents the location for the $l^{th}$ regime corresponding to $(x_m)_i$, and the truncation error is $O(k^2 + h^4)$. The optimal exercise boundary and the initial and boundary conditions for each regime are calculated as

$$s_{f(m)}^{n+1} = K - (u_m)_0^{n+1}, \quad (w_m)_0^{n+1} = -s_{f(m)}^{n+1}, \quad (y_m)_0^{n+1} = -s_{f(m)}^{n+1}, \quad (z_m)_0^{n+1} = -s_{f(m)}^{n+1}\,; \quad (21)$$

$$(u_m)_M^{n+1} = 0, \quad (w_m)_M^{n+1} = 0, \quad (y_m)_M^{n+1} = 0, \quad (z_m)_M^{n+1} = 0; \quad (22)$$

$$(u_m)_i^0 = (w_m)_i^0 = (y_m)_i^0 = (z_m)_i^0 = 0, \qquad i = 1,2,\cdots,M. \quad (23)$$

Let the approximate solutions of the theta, delta decay, and color options for each regime be given as

$$\frac{\partial U_m((x_m)_i,\, \tau_n)}{\partial \tau} \approx (\Theta_m)_i^n, \qquad \frac{\partial W_m((x_m)_i,\, \tau_n)}{\partial \tau} \approx (\mathrm{K}_m)_i^n, \qquad \frac{\partial Y_m((x_m)_i,\, \tau_n)}{\partial \tau} \approx (\Gamma_m)_i^n, \quad (24)$$

respectively. For $n = 1$, we approximate the three Greeks using the first-order backward finite differences

$$(\Theta_m)_i^1 \approx \frac{(u_m)_i^1 - (u_m)_i^0}{k}, \qquad (\mathrm{K}_m)_i^1 \approx \frac{(w_m)_i^1 - (w_m)_i^0}{k}, \qquad (\Gamma_m)_i^1 \approx \frac{(y_m)_i^1 - (y)_i^0}{k}. \quad (25a)$$

Subsequently, we use the second-order backward finite difference approximations as

$$(\Theta_m)_i^{n+1} = \frac{3(u_m)_i^{n+1} - 4(u_m)_i^n + (u_m)_i^{n-1}}{2k}, \qquad (\mathrm{K}_m)_i^{n+1} = \frac{3(w)_i^{n+1} - 4(w_m)_i^n + (w_m)_i^{n-1}}{2k}; \tag{25b}$$

$$(\Gamma_m)_i^{n+1} = \frac{3(y)_i^{n+1} - 4(y_m)_i^n + (y_m)_i^{n-1}}{2k}. \tag{25c}$$

The initial conditions of the theta, delta decay, and color options for each regime are calculated as

$$(\Theta_m)_i^0 = 0, \qquad (\mathrm{K}_m)_i^0 = 0, \qquad (\Gamma_m)_i^0 = 0, \qquad i = 0,1,\cdots,M. \tag{26}$$

### 3.2. Hermite Interpolation

Note that the relationship between the fixed interval (and the mesh) for the $l^{th}$ regime and the fixed interval (and the mesh) for the $m^{th}$ regime after the logarithmic transformation is as follows:

$$(x_l)_j = (x_m)_i - \ln\frac{s_{f(l)}(\tau_n)}{s_{f(m)}(\tau_n)}. \tag{27}$$

If $s_{f(l)}(\tau_n) = s_{f(m)}(\tau_n)$, then the fixed interval for the $l^{th}$ regime overlaps completely with the fixed interval for the $m^{th}$ regime. Hence, $(x_m)_i = (x_l)_j$ from (27) and we solve only one interval. However, if $s_{f(l)}(\tau_n) \neq s_{f(m)}(\tau_n)$, we need to consider three cases as shown in Fig. 1 and (28).

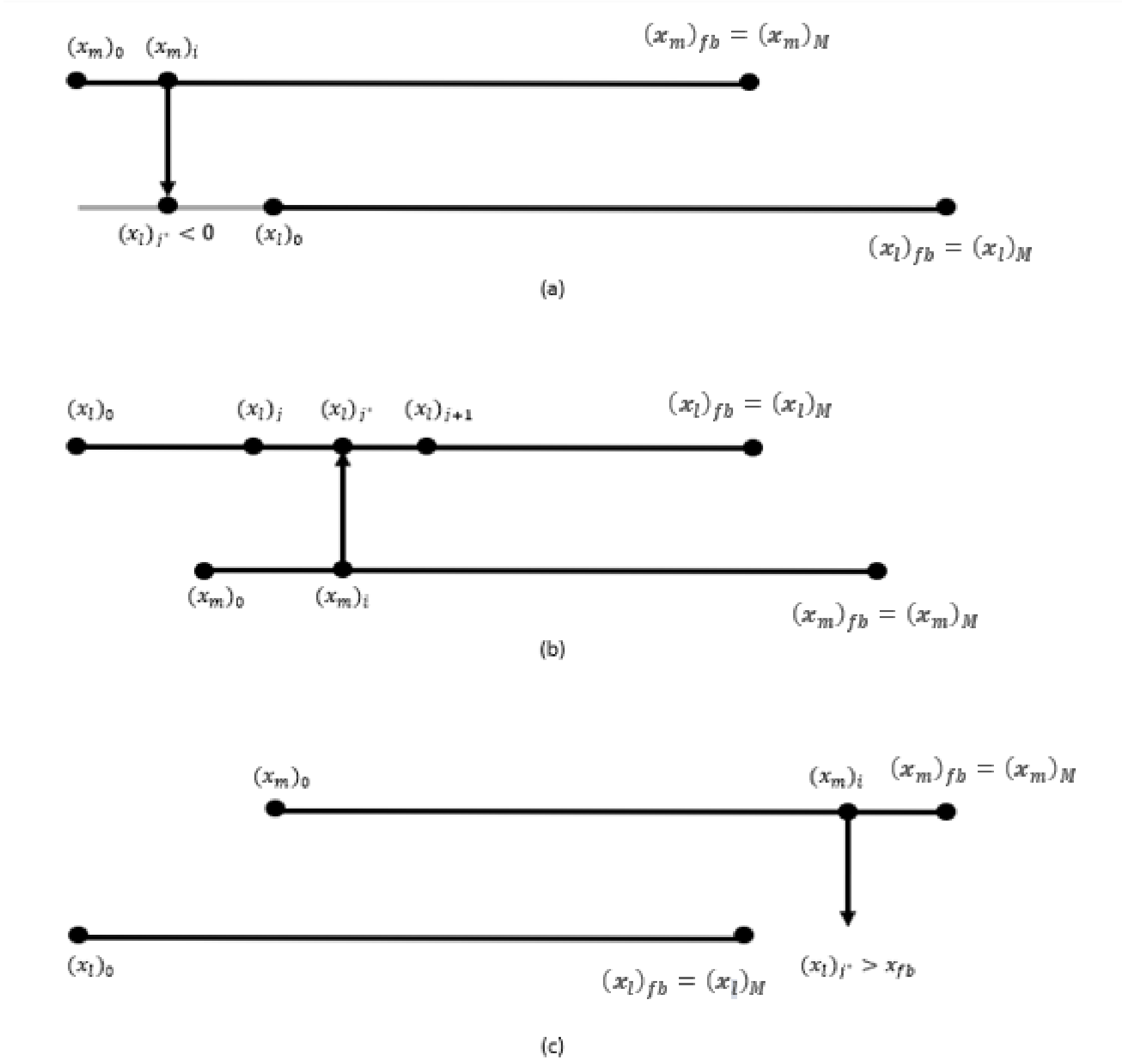

**Fig. 1**. Relationship between the $l^{th}$ and $m^{th}$ intervals and the location of the $(x_m)_i$ in the $l^{th}$ interval.

$$(x_l)_{j^*} = (x_m)_i - \ln\frac{s_{f(l)}(\tau_n)}{s_{f(m)}(\tau_n)} < 0, \tag{28a}$$

$$(x_l)_{j^*} = (x_m)_i - \ln\frac{s_{f(l)}(\tau_n)}{s_{f(m)}(\tau_n)} > (x_m)_M, \tag{28b}$$

$$(x_l)_j < (x_l)_{j^*} = (x_m)_i - \ln\frac{s_{f(l)}(\tau_n)}{s_{f(m)}(\tau_n)} < (x_l)_{j+1}. \tag{28c}$$

For (28a), we obtain the value of the asset option and its derivatives as follows:

$$(u_l)_{j^*}^n = K - e^{(x_l)_{j^*}} s_{f(l)}, (w_l)_{j^*}^n = (y_l)_{j^*}^n = (z_l)_{j^*}^n = -e^{(x_l)_{j^*}} s_{f(l)}. \tag{28d}$$

For (28b),

$$(u_l)_{j^*}^n = (w_l)_{j^*}^n = (y_l)_{j^*}^n = (z_l)_{j^*}^n = 0. \tag{28e}$$

Evaluation of (28c) requires interpolating between nodes. Since $(x_l)_{j^*} \in [(x_l)_j, (x_l)_{j+1}]$, $(u_l)_{j^*}^n, (w_l)_{j^*}^n, (y_l)_{j^*}^n$ and $(z_l)_{j^*}^n$ need to be evaluated using an interpolation on $(u_l)_j^n, (w_l)_j^n, (y_l)_j^n$ and $(z_l)_j^n$. We employ the Hermite interpolation with a Newton basis [35,11,6] to estimate these coupled regimes in the set of equations.

Let $j^* \in [j, j+1]$ be the point in the interval of $l^{th}$ regime that we need to approximate the asset, delta, gamma, and speed function. Hence,

$$(u_l)_{j^*}^n = \alpha_{u,o} + \alpha_{u,1}[(x_l)_{j^*} - (x_l)_j] + \alpha_{u,2}[(x_l)_{j^*} - (x_l)_j]^2 + \alpha_{u,3}[(x_l)_{j^*} - (x_l)_j]^2[(x_l)_{j^*} - (x_l)_{j+1}], \tag{29a}$$

$$(w_l)_{j^*}^n = \alpha_{u,1} + 2\alpha_{u,2}[(x_l)_{j^*} - (x_l)_j] + 2\alpha_{u,3}[(x_l)_{j^*} - (x_l)_j][(x_l)_{j^*} - (x_l)_{j+1}] + \alpha_{u,3}[(x_l)_{j^*} - (x_l)_j]^2, \tag{29b}$$

$$(y_l)_{j^*}^n = \alpha_{y,o} + \alpha_{y,1}[(x_l)_{j^*} - (x_l)_j] + \alpha_{y,2}[(x_l)_{j^*} - (x_l)_j]^2 + \alpha_{y,3}[(x_l)_{j^*} - (x_l)_j]^2[(x_l)_{j^*} - (x_l)_{j+1}], \tag{29c}$$

$$(z_l)_{j^*}^n = \alpha_{z,o} + \alpha_{z,1}[(x_l)_{j^*} - (x_l)_j] + \alpha_{z,2}[(x_l)_{j^*} - (x_l)_j]^2 + \alpha_{z,3}[(x_l)_{j^*} - (x_l)_j]^2[(x_l)_{j^*} - (x_l)_{j+1}], \tag{29d}$$

Here,

$$\alpha_{u,o} = (u_l)_j^n, \quad \alpha_{u,1} = (w_l)_j^n, \quad \alpha_{u,2} = \frac{1}{h}\left(\frac{[(u_l)_{j+1}^n - (u_l)_j^n]}{h} - \alpha_{u,1}\right); \tag{29e}$$

$$\alpha_{u,3} = \frac{1}{h}\left[\frac{1}{h}\left((w_l)_{j+1}^n - \frac{[(u_l)_{j+1}^n - (u_l)_j^n]}{h}\right) - \alpha_{u,2}\right], \tag{29f}$$

$$\vdots$$

$$\alpha_{z,o} = (z_l)_j^n, \quad \alpha_{z,1} = (\hat{z}_l)_j^n, \quad \alpha_{z,2} = \frac{1}{h}\left(\frac{[(z_l)_{j+1}^n - (z_l)_j^n]}{h} - \alpha_{z,1}\right); \tag{29g}$$

$$\alpha_{z,3} = \frac{1}{h}\left[\frac{1}{h}\left((\hat{z}_l)_{j+1}^n - \frac{[(z_l)_{j+1}^n - (z_l)_j^n]}{h}\right) - \alpha_{z,2}\right] \tag{29h}$$

**Remark 4**: It is worth noting that the derivative of the speed option, $(\hat{z}_l)_j^n$, is employed in the cubic Hermite interpolations. This is because interpolating gamma and speed options using the derivative of the Hermite function result in non-smooth solutions. Hence, we interpolate the gamma and speed option directly from the Hermite function which results in the use of $(\hat{z}_l)_j^n$ for such an evaluation. An alternative could have been to use other interpolation methods that do not involve derivatives of a function, however, Hermite interpolation proves to be very fast and more efficient for our model. We approximate $(\hat{z}_l)_j^n$ with fourth-order accuracy by further taking derivative of (5d) and discretize with compact and Crank Nicholson scheme in space and time, respectively.

## 4. Multigrid Method

In the previous method, we used the standalone Gauss-Seidel iteration to approximate the numerical solutions of the optimal exercise boundary, asset option, and option Greeks for each regime. However, we found that as the grid sizes $k$ and $h$ decrease, the convergence of our numerical solutions becomes very slow. Moreover, for each iteration, the coupled regime with high-order interpolation further increases the computational burden.

Thus, in this section, we will employ the multigrid method to improve computation. To use the multigrid method, we first start our iteration on a fine uniform grid

$$A_{m,h}(\boldsymbol{u}_m)_{\boldsymbol{h}} = \boldsymbol{f}_{m,h}^u, \qquad B_{m,h}(\boldsymbol{w}_m)_{\boldsymbol{h}} = \boldsymbol{f}_{m,h}^w, \qquad B_{m,h}(\boldsymbol{y}_m)_{\boldsymbol{h}} = \boldsymbol{f}_{m,h}^y, \qquad B_{m,h}(\boldsymbol{z}_m)_{\boldsymbol{h}} = \boldsymbol{f}_{m,h}^z, \tag{30}$$

where $h$ is the step size of the finest grid. Next, we relax (30) $v_1$ times using an iterative method that is a good smoother and computes their residuals as follows:

$$\boldsymbol{r}_{m,h}^u = \boldsymbol{f}_{m,h}^u - A_{m,h}(\boldsymbol{u}_m)_{\boldsymbol{h}}, \qquad \boldsymbol{r}_{m,h}^w = \boldsymbol{f}_{m,h}^w - B_{m,h}(\boldsymbol{w}_m)_{\boldsymbol{h}}, \tag{31a}$$

$$\boldsymbol{r}_{m,h}^y = \boldsymbol{f}_{m,h}^y - B_{m,h}(\boldsymbol{y}_m)_{\boldsymbol{h}}, \qquad \boldsymbol{r}_{m,h}^z = \boldsymbol{f}_{m,h}^z - B_{m,h}(\boldsymbol{z}_m)_{\boldsymbol{h}}. \tag{31b}$$

After computing the residuals, we transfer the latter to a coarse grid using the restriction operator

$$\boldsymbol{r}_{m,2h}^u = \mathcal{R}_h^{2h}\boldsymbol{r}_{m,h}^u, \qquad \boldsymbol{r}_{m,2h}^w = \mathcal{R}_h^{2h}\boldsymbol{r}_{m,h}^w, \qquad \boldsymbol{r}_{m,2h}^y = \mathcal{R}_h^{2h}\boldsymbol{r}_{m,h}^y, \qquad \boldsymbol{r}_{m,2h}^z = \mathcal{R}_h^{2h}\boldsymbol{r}_{m,h}^z. \tag{32}$$

The restriction operator can be calculated from the prolongation matrix as follows:

$$\mathcal{P}_{2h}^{h} = \frac{1}{2}\begin{bmatrix} 2 & & & & \\ 1 & 1 & & & \\ & 2 & & & \\ & \ddots & \ddots & & \\ & 1 & 1 & & \\ & & 2 & & \\ & & 1 & 1 \\ & & & 2 \end{bmatrix}, \qquad \mathcal{R}_{h}^{2h} = \frac{1}{2}\left(\mathcal{P}_{2h}^{h}\right)^{T}. \tag{33}$$

To avoid matrix-vector multiplication and properly include the Neumann boundary effect on the error and the residual for the asset options, we employ the method described in the work of Briggs et. al [5] and obtain $\boldsymbol{r}_{m,2h}^{u}$ directly from $\boldsymbol{r}_{m,h}^{u}$ as follows:

$$\left(\boldsymbol{r}_{m,2h}^{u}\right)_{i}^{n+1} = \frac{\left(\boldsymbol{r}_{m,h}^{u}\right)_{2i-1}^{n+1} + 2\left(\boldsymbol{r}_{m,h}^{u}\right)_{2i}^{n+1} + \left(\boldsymbol{r}_{m,h}^{u}\right)_{2i+1}^{n+1}}{4}, \tag{34a}$$

$$\vdots$$

$$\left(\boldsymbol{r}_{m,2h}^{z}\right)_{i}^{n+1} = \frac{\left(\boldsymbol{r}_{m,h}^{z}\right)_{2i-1}^{n+1} + 2\left(\boldsymbol{r}_{m,h}^{z}\right)_{2i}^{n+1} + \left(\boldsymbol{r}_{m,h}^{z}\right)_{2i+1}^{n+1}}{4}, \qquad i = 1,2,\dots,(M/2)-1, \tag{34b}$$

$$\left(\boldsymbol{r}_{m,2h}^{u}\right)_{0}^{n+1} = \frac{2\left(\boldsymbol{r}_{m,h}^{u}\right)_{0}^{n+1} + \left(\boldsymbol{r}_{m,h}^{u}\right)_{1}^{n+1}}{4}, \tag{35a}$$

$$\left(\boldsymbol{r}_{m,2h}^{w}\right)_{0}^{n+1} = \left(\boldsymbol{r}_{m,h}^{w}\right)_{0}^{n+1}, \qquad \left(\boldsymbol{r}_{m,2h}^{y}\right)_{0}^{n+1} = \left(\boldsymbol{r}_{m,h}^{y}\right)_{0}^{n+1}, \qquad \left(\boldsymbol{r}_{m,2h}^{Z}\right)_{0}^{n+1} = \left(\boldsymbol{r}_{m,h}^{Z}\right)_{0}^{n+1}; \tag{35b}$$

$$\left(\boldsymbol{r}_{m,2h}^{u}\right)_{M/2}^{n+1} = \left(\boldsymbol{r}_{m,h}^{u}\right)_{M}^{n+1}, \qquad \left(\boldsymbol{r}_{m,2h}^{w}\right)_{M/2}^{n+1} = \left(\boldsymbol{r}_{m,h}^{w}\right)_{M}^{n+1}; \tag{36a}$$

$$\left(\boldsymbol{r}_{m,2h}^{y}\right)_{M/2}^{n+1} = \left(\boldsymbol{r}_{m,h}^{y}\right)_{M}^{n+1}, \qquad \left(\boldsymbol{r}_{m,2h}^{Z}\right)_{M/2}^{n+1} = \left(\boldsymbol{r}_{m,h}^{Z}\right)_{M}^{n+1}. \tag{36b}$$

Next, we obtain the exact solution of the defect equations

$$A_{m,2h}\boldsymbol{e}_{m,2h}^{u} = \boldsymbol{r}_{m,2h}^{u}, \qquad B_{m,2h}\boldsymbol{e}_{m,2h}^{w} = \boldsymbol{r}_{m,2h}^{w}, \qquad B_{m,2h}\boldsymbol{e}_{m,2h}^{y} = \boldsymbol{r}_{m,2h}^{y}, \qquad B_{m,2h}\boldsymbol{e}_{m,2h}^{z} = \boldsymbol{r}_{m,2h}^{z}, \tag{37}$$

where $\boldsymbol{e}_{2h}$ is the error term on the coarse grid. We then transfer the error term to the fine grid using prolongation (interpolation) operators

$$\boldsymbol{e}_{m,h}^{u} = \mathcal{P}_{2h}^{h}\boldsymbol{e}_{m,2h}^{u}, \qquad \boldsymbol{e}_{m,h}^{w} = \mathcal{P}_{2h}^{h}\boldsymbol{e}_{m,2h}^{w}, \qquad \boldsymbol{e}_{m,h}^{y} = \mathcal{P}_{2h}^{h}\boldsymbol{e}_{m,2h}^{y}, \qquad \boldsymbol{e}_{m,h}^{z} = \mathcal{P}_{2h}^{h}\boldsymbol{e}_{m,2h}^{z}. \tag{38}$$

Finally, we correct our approximation of the asset, delta, gamma, and speed options

$$(\boldsymbol{u}_m)_{\boldsymbol{h}} = (\boldsymbol{u}_m)_{\boldsymbol{h}} + \boldsymbol{e}_{m,h}^{u}, \qquad (\boldsymbol{w}_m)_{\boldsymbol{h}} = (\boldsymbol{w}_m)_{\boldsymbol{h}} + \boldsymbol{e}_{m,h}^{w}, \tag{39a}$$

$$(\boldsymbol{y}_m)_{\boldsymbol{h}} = (\boldsymbol{y}_m)_{\boldsymbol{h}} + \boldsymbol{e}_{m,h}^{y}, \qquad (\boldsymbol{z}_m)_{\boldsymbol{h}} = (\boldsymbol{z}_m)_{\boldsymbol{h}} + \boldsymbol{e}_{m,h}^{z}. \tag{39b}$$

and relax (30) $\upsilon_2$ times using a smoothing method with our new approximation as the initial guess.

### 4.1. Smoothing Method

Good smoothers are components of effective multigrid methods. This is because few iterations with those relaxation methods remove the high-frequency error. In this work, we use the Gauss-Seidel iteration as the smoother by rearranging the compact scheme in (14), (16) in a pointwise manner as follows:

$$(u_m^{k+1})_0^{n+1} = \frac{\left(f_m^{u^k}\right)_0^{n+1/2} - b_m^1 (u_m^k)_1^{n+1}}{a_m^1}, \tag{40a}$$

$$(u_m^{k+1})_i^{n+1} = \frac{\left(f_m^{u^k}\right)_i^{n+1/2} - d_m^1 (u_m^{k+1})_{i-1}^{n+1} - d_m^1 (u_m^k)_{i+1}^{n+1}}{c_m^1}, \tag{40b}$$

$$(w_m^{k+1})_i^{n+1} = \frac{\left(f_m^{w^k}\right)_i^{n+1/2} - d_m^1 (w_m^{k+1})_{i-1}^{n+1} - d_m^1 (w_m^k)_{i+1}^{n+1}}{c_m^1}, \tag{41}$$

$$(y_m^{k+1})_i^{n+1} = \frac{\left(f_m^{y^k}\right)_i^{n+1/2} - d_m^1 (y_m^{k+1})_{i-1}^{n+1} - d_m^1 (y_m^k)_{i+1}^{n+1}}{c_m^1}, \tag{42}$$

$$(z_m^{k+1})_i^{n+1} = \frac{\left(f_m^{z^k}\right)_i^{n+1/2} - d_m^1 (z_m^{k+1})_{i-1}^{n+1} - d_m^1 (z_m^k)_{i+1}^{n+1}}{c_m^1}, \qquad i = 1,2,\cdots,M-1. \tag{43}$$

### 4.2. Multigrid Algorithm

Here, we implement a modified multigrid strategy presented in the work of Hafner and Konke [17]. They mentioned that it is necessary for balancing computational effort in each mesh. In this method, rather than recursive restriction and smoothing across each coarse grid, it is done once on a coarse grid and interpolated back to the fine grid in an increasing fashion starting from the coarsest grid. We observe during the numerical experiment that interpolating with cubic or high order interpolation speeds up convergence with this method. Another distinctness of this approach is that rather than solving the defect equation in exact form, a smoothing process is carried out with many iteration steps. The number of inner iteration steps in each interior smoothing is given as follow:

$$s_i = c(q-i)^2, \qquad i = 0, 1, \dots, q-1 \tag{44}$$

where $c$ is an arbitrary constant factor. $i = 0$ and $i = q$ represent the coarsest and the finest grid, respectively. An algorithm for obtaining the numerical solutions of the optimal exercise boundary, asset option, and the option Greeks in each regime using the M-cycle multigrid method is described below.

**Algorithm 2.** An algorithm based on the Modified-cycle multigrid (M-cycle)

1. Initialize $s^n_{f(m)}$, $(\boldsymbol{u}_m)^n_h$, $(\boldsymbol{w}_m)^n_h$, $(\boldsymbol{y}_m)^n_h$, $(\boldsymbol{z}_m)^n_h$, $(\boldsymbol{\Theta}_m)^n_h$, $(\mathbf{K}_m)^n_h$, and $(\boldsymbol{\Gamma}_m)^n_h$ for $m = 1,2,\dots,I$.
2. **for n = 1 to N**
3. Compute $(\boldsymbol{u}_l)^n_h$, $(\boldsymbol{w}_l)^n_h$, $(\boldsymbol{y}_l)^n_h$, and $(\boldsymbol{z}_l)^n_h$ for $l = 1,2,\dots,I$ and $l \neq m$ based on (29)
4. $Set\ s^{n+1(it=0)}_{f(m)} = s^n_{f(m)}$, $(\boldsymbol{u}_m)^{n+1(it=0)}_h = (\boldsymbol{u}_m)^n_h$, $(\boldsymbol{w}_m)^{n+1(it=0)}_h = (\boldsymbol{w}_m)^n_h$, $(\boldsymbol{y}_m)^{n+1(it=0)}_h = (\boldsymbol{y}_m)^n_h$, and $(\boldsymbol{z}_m)^{n+1(it=0)}_h = (\boldsymbol{z}_m)^n_h$.
5. **while true**
6. Compute $(\boldsymbol{u}_l)^{n+1(it)}_h$, …, and $(\boldsymbol{z}_l)^{n+1(it)}_h$ for $l = 1,2,\dots,I$ and $l \neq m$ based on (29)
7. Relax $A_{m,h}(\boldsymbol{u}_m)^{n+1(it+1/n)}_h = \boldsymbol{f}^u_{m,h}$, …, and $B_{m,h}(\boldsymbol{z}_m)^{n+1(it+1/n)}_h = \boldsymbol{f}^z_{m,h}$, and $s^{n+1(it+1/n)}_{f(m)}$ $\upsilon_1$ times using either on (40)-(43) with the initial guess $(\boldsymbol{u}_m)^{n+1(it)}_h$, …, and $(\boldsymbol{z}_m)^{n+1(it)}_h$, and $s^{n+1(it)}_{f(m)}$
8. Compute $\boldsymbol{r}^u_{m,h}$, …, and $\boldsymbol{r}^z_{m,h}$ based on (31).
9. **If** $\max\limits_{1\le m\le I}\left|s^{n+1(\mathrm{It}+1/n)}_{f(m)} - s^{n+1(\mathrm{It}+1/n)}_{f(m)}\right| < \varepsilon$ and $\max\limits_{1\le m\le I}\left|\boldsymbol{r}^u_{m,h}\right| < \varepsilon$
10. Set $(\boldsymbol{u}_m)^n_h = (\boldsymbol{u}_m)^{n+1}_h$, …, and $(\boldsymbol{z}_m)^n_h = (\boldsymbol{z}_m)^{n+1}_h$, and $s^n_{f(m)} = s^{n+1}_{f(m)}$. Compute $(\boldsymbol{\Theta}_m)^{n+1}_h$, $(\mathbf{K}_m)^{n+1}_h$, and $(\boldsymbol{\Gamma}_m)^{n+1}_h$. **break**
11. **else**, go to 10
12. For $i = n-1, n-2, \dots, 1$, compute recursively:
13. Compute $\boldsymbol{r}^u_{m,2^ih}, \dots, \boldsymbol{r}^z_{m,2^ih}$
14. Relax $A_{m,2^ih}\boldsymbol{e}^u_{m,2^ih} = \boldsymbol{r}^u_{m,2^ih}$, …, and $B_{m,2^ih}\boldsymbol{e}^z_{m,2^ih} = \boldsymbol{r}^z_{m,2^ih}$ with an initial guess of $\boldsymbol{e}^u_{m,2^ih} = \cdots = \boldsymbol{e}^z_{m,2^ih} = 0$
15. Compute $\boldsymbol{e}^u_{m,h}$,…, $\boldsymbol{e}^z_{m,h}$ from $\boldsymbol{e}^u_{m,2^ih}, \dots, \boldsymbol{e}^z_{m,2^ih}$ with cubic or high order interpolation.
16. Compute $(\boldsymbol{u}_m)^{n+1[it+(n-i+1/n)]}_h = (\boldsymbol{u}_m)^{n+1[it+(n-i/n)]}_h + \boldsymbol{e}^u_{m,h}$ , …, and $(\boldsymbol{z}_m)^{n+1[it+(n-i/n)+1]}_h = (\boldsymbol{z}_m)^{n+1[it+(n-i/n)]}_h + \boldsymbol{e}^z_{m,h}$ based on (39)
17. Compute $(\boldsymbol{u}_l)^{n+1}_h$, $(\boldsymbol{w}_l)^{n+1}_h$, $(\boldsymbol{y}_l)^{n+1}_h$, and $(\boldsymbol{z}_l)^{n+1}_h$ for $l = 1,2,\dots,I$ and $l \neq m$ based on (29)
18. Relax $A_{m,h}(\boldsymbol{u}_m)^{n+1[it+(n-i+1/n)]}_h = \boldsymbol{f}^u_{m,h}$, …, and $B_{m,h}(\boldsymbol{z}_m)^{n+1[it+(n-i+1/n)]}_h = \boldsymbol{f}^z_{m,h}$ and $s^{n+1[it+(n-i+1/n)]}_{f(m)}$ $\upsilon_2$ times based on (40)-(43)
19. **endif**
20. **endwhile**
21. **endfor**

Furthermore, to check the effect of good initial approximation on the convergence of our numerical solution, we implemented our M-cycle multigrid strategy with full multigrid initialization. It is well known that the full multigrid method can be used for solving linear and nonlinear PDEs by providing a good initial guess. The first phase involves using the full multigrid method to obtain a good initial approximation. It is important to mention that we used cubic interpolation for both grid correction and obtaining an approximate initial solution at the finer grid. The second phase involves implementing the M-cycle with the initial approximation solutions obtained from the full multigrid method.

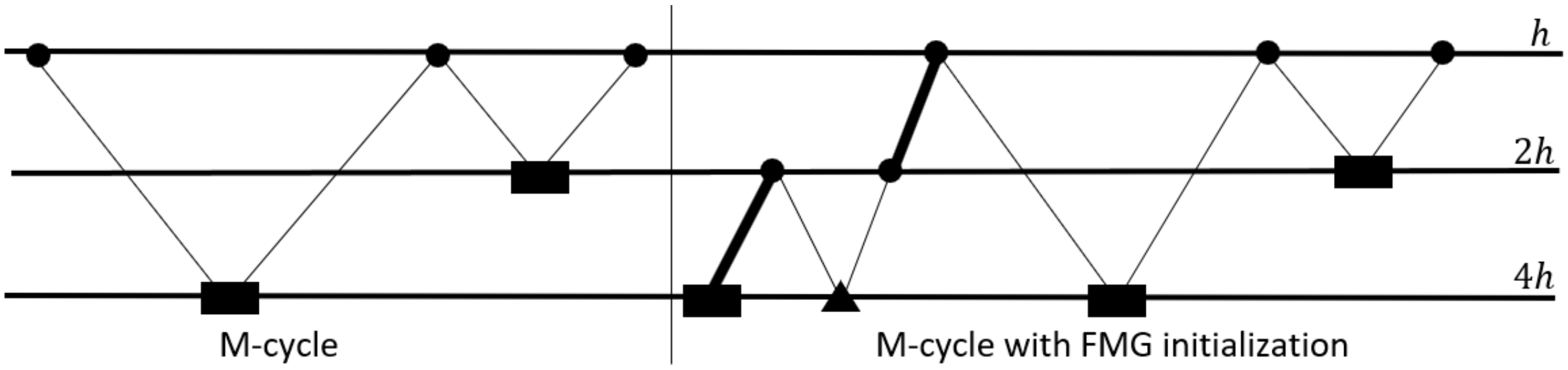

**Fig. 2.** Three grid sequence modified multigrid strategy with a large number of smoothing ■, few numbers of smoothing ●, high order interpolation ⁄, and exact computation ▲.

## 5. Numerical Experiment

To check the performance of our proposed algorithm, we test its accuracy in a couple of numerical examples where we consider American put options with two- and four-regime. The numerical experiment was carried out on the mesh with a uniform grid size. The numerical code was written with MATLAB 2019a on Intel Core i5-3317U CPU 1.70GHz 64-bit ASUS Laptop.

### 5.1. Two-Regime Example

**Example 1:** Consider a two-regime problem with the following data:

$$K = 9, \qquad T = 1, \qquad Q = \begin{bmatrix} -6 & 6 \\ 9 & -9 \end{bmatrix}, \qquad \boldsymbol{r} = \begin{bmatrix} 0.10 \\ 0.05 \end{bmatrix}, \qquad \boldsymbol{\sigma} = \begin{bmatrix} 0.80 \\ 0.30 \end{bmatrix}, \qquad \varepsilon = 10^{-8}. \tag{45}$$

In our computation, we chose the interval $0 \leq x_m \leq 3$. Our time step $k$ was chosen such that $k = h^2$. We label results based on the multigrid method as follows:

- $M(i, \cdots, i)$ – M-cycle with $i$ outer Gauss-Seidel smoothing.
- $M_{FMG}(i, \cdots, i)$ – M-cycle with full multigrid initialization and $i$ outer Gauss-Seidel smoothing.

To validate the accuracy of our multigrid method, we compared our results with MTree [29] IMS1, IMS2 [23], and MOL [7]. We listed the results in Tables 1 and 2. The plots of the option price, option Greeks, and optimal exercise boundary for each regime were displayed in Fig 3. Furthermore, we computed the log scale of the absolute error and normalized residual in the sense of $l_2$ norm per each iteration at the final time level and displayed the plot in Fig. 4. From Tables 1-2, based on the comparison between the multigrid method and other existing methods, the results from the multigrid method are very close to MTree and MOL. From Fig. 4, the error per iteration reduces faster in the multigrid method when compared with our previous method.

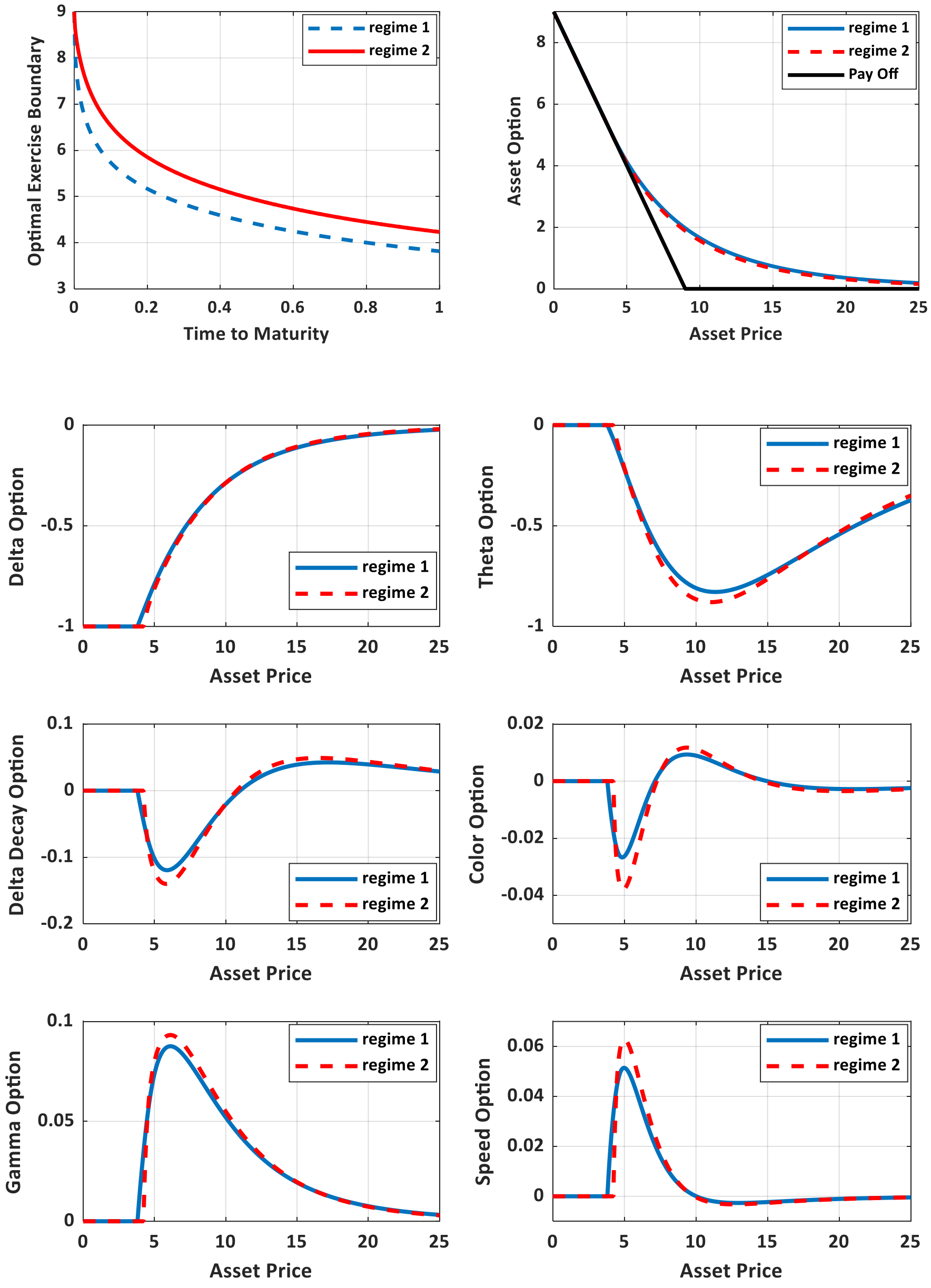

**Fig. 3.** Optimal exercise boundaries, asset options, and option Greeks for example 1 ($\tau = T$).

We verified the global maximum iteration and the average CPU time of our previous method with the multigrid method and displayed the results in Table 3. From Table 3, with a tolerance of $\varepsilon = 10^{-8}$, the global maximum number of iterations and the CPU time(s) required to achieve numerical convergence is smaller in the multigrid method when compared with the classical Gauss-Seidel iteration. Furthermore, based on the three grid sequences from Table 3, the M-cycle with full multigrid initialization has less CPU time when compared with the conventional M-cycle multigrid method.

**Table 1.** Comparison of the American put option price in regime 1.

| S | MTree | IMS1 | IMS2 | MOL | M-cycle | |
|---|---|---|---|---|---|---|
| | | | | | $h = 0.0125$ | $0.00625$ |
| 3.5 | 5.5000 | 5.5001 | 5.5001 | 5.5000 | 5.5000 | 5.5000 |
| 4.0 | 5.0031 | 5.0067 | 5.0066 | 5.0033 | 5.0033 | 5.0033 |
| 4.5 | 4.5432 | 4.5486 | 4.5482 | 4.5433 | 4.5433 | 4.5433 |
| 6.0 | 3.4144 | 3.4198 | 3.4184 | 3.4143 | 4.4143 | 3.4143 |
| 7.5 | 2.5844 | 2.5877 | 2.5867 | 2.5842 | 2.5842 | 2.5842 |
| 8.5 | 2.1560 | 2.1598 | 2.1574 | 2.1559 | 2.1559 | 2.1558 |
| 9.0 | 1.9722 | 1.9756 | 1.9731 | 1.9720 | 1.9720 | 1.9720 |
| 9.5 | 1.8058 | 1.8090 | 1.8064 | 1.8056 | 1.8056 | 1.8056 |
| 10.5 | 1.5186 | 1.5214 | 1.5187 | 1.5185 | 1.5185 | 1.5185 |
| 12.0 | 1.1803 | 1.1827 | 1.1799 | 1.1803 | 1.1804 | 1.1803 |

**Table 2.** Comparison of the American put option price in regime 2.

| S | MTree | IMS1 | IMS2 | MOL | M-cycle | |
|---|---|---|---|---|---|---|
| | | | | | $h = 0.0125$ | $0.00625$ |
| 3.5 | 5.5000 | 5.5012 | 5.5012 | 5.5000 | 5.5000 | 5.5000 |
| 4.0 | 5.0000 | 5.0016 | 5.0016 | 5.0000 | 5.0000 | 5.0000 |
| 4.5 | 4.5117 | 4.5194 | 4.5190 | 4.5119 | 4.5120 | 4.5119 |
| 6.0 | 3.3503 | 3.3565 | 3.3550 | 3.3507 | 3.3507 | 3.3507 |
| 7.5 | 2.5028 | 2.5078 | 2.5056 | 2.5033 | 2.5034 | 2.5033 |
| 8.5 | 2.0678 | 2.0722 | 2.0695 | 2.0683 | 2.0684 | 2.0683 |
| 9.0 | 1.8819 | 1.8860 | 1.8832 | 1.8825 | 1.8825 | 1.8824 |
| 9.5 | 1.7143 | 1.7181 | 1.7153 | 1.7149 | 1.7150 | 1.7149 |
| 10.5 | 1.4267 | 1.4301 | 1.4272 | 1.4273 | 1.4274 | 1.4273 |
| 12.0 | 1.0916 | 1.0945 | 1.0916 | 1.0923 | 1.0924 | 1.0923 |

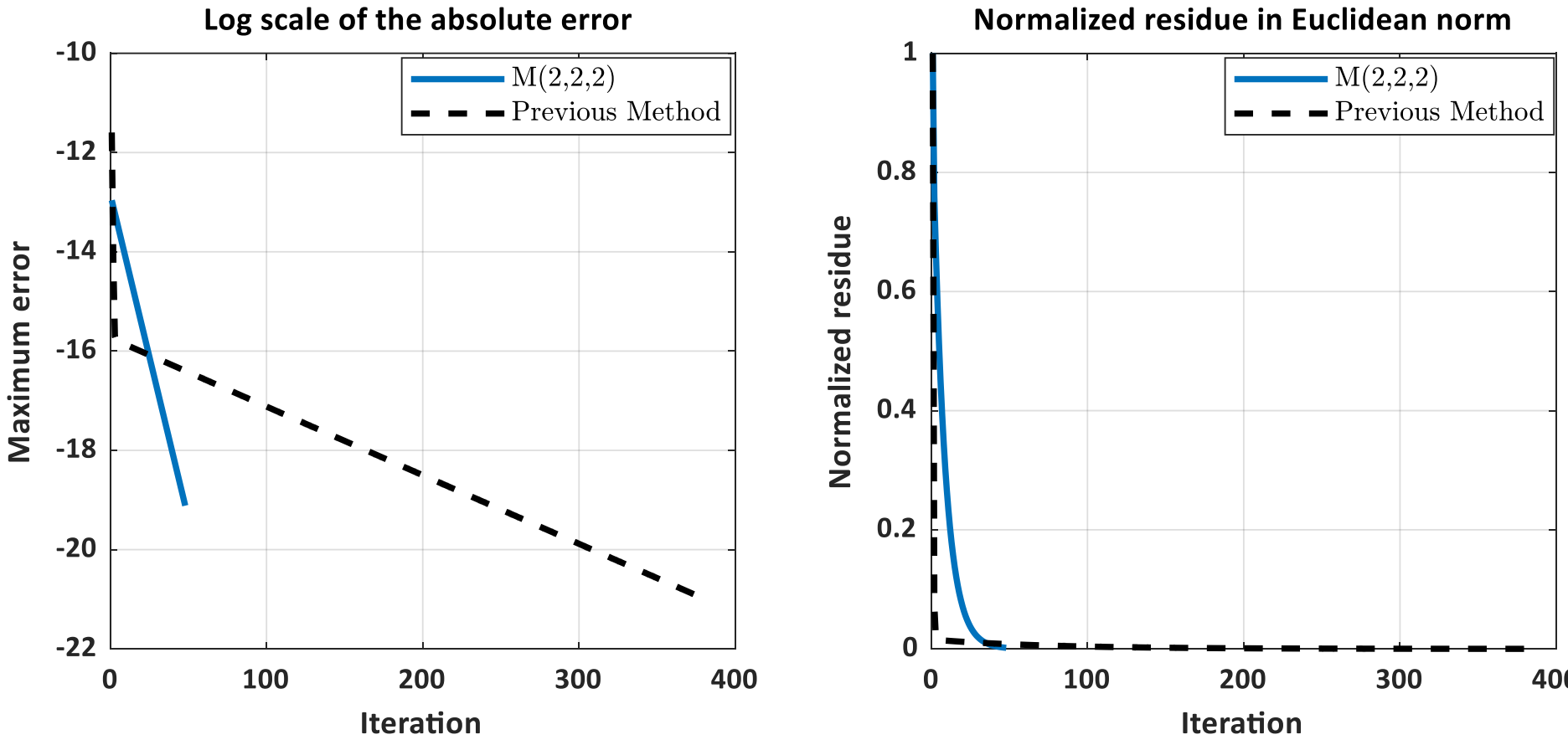

**Fig. 4.** $l_2$ norm of the normalized residue and log scale of the absolute error in Regime 1 ($c = 3$).

Next, we computed the convergent factor of the multigrid method. For the numerical convergent factor at each time level, we estimated it as follows [43,14]:

$$\varrho = \sqrt[m]{\frac{\|\boldsymbol{r}^m\|_2}{\|\boldsymbol{r}^0\|_2}}, \tag{46}$$

where $\boldsymbol{r}^m$ is the residual at $m$ iteration for a given time level. Tronttenberg et al. [43] mentioned that if $m$ is sufficiently large, (46) is a good estimate of the convergent factor. The result of the measured convergent factor was displayed in Fig. 5. In Table 4, we listed the measured convergent factors based on variable step sizes and fixed a time step $k = 0.0001$. We used the measured convergent factor to verify the $h$ -independency of the M-cycle multigrid. For a fixed time step, we observed as listed in Table 4 that the measured convergent factor increases as $k/h^2$ increases.

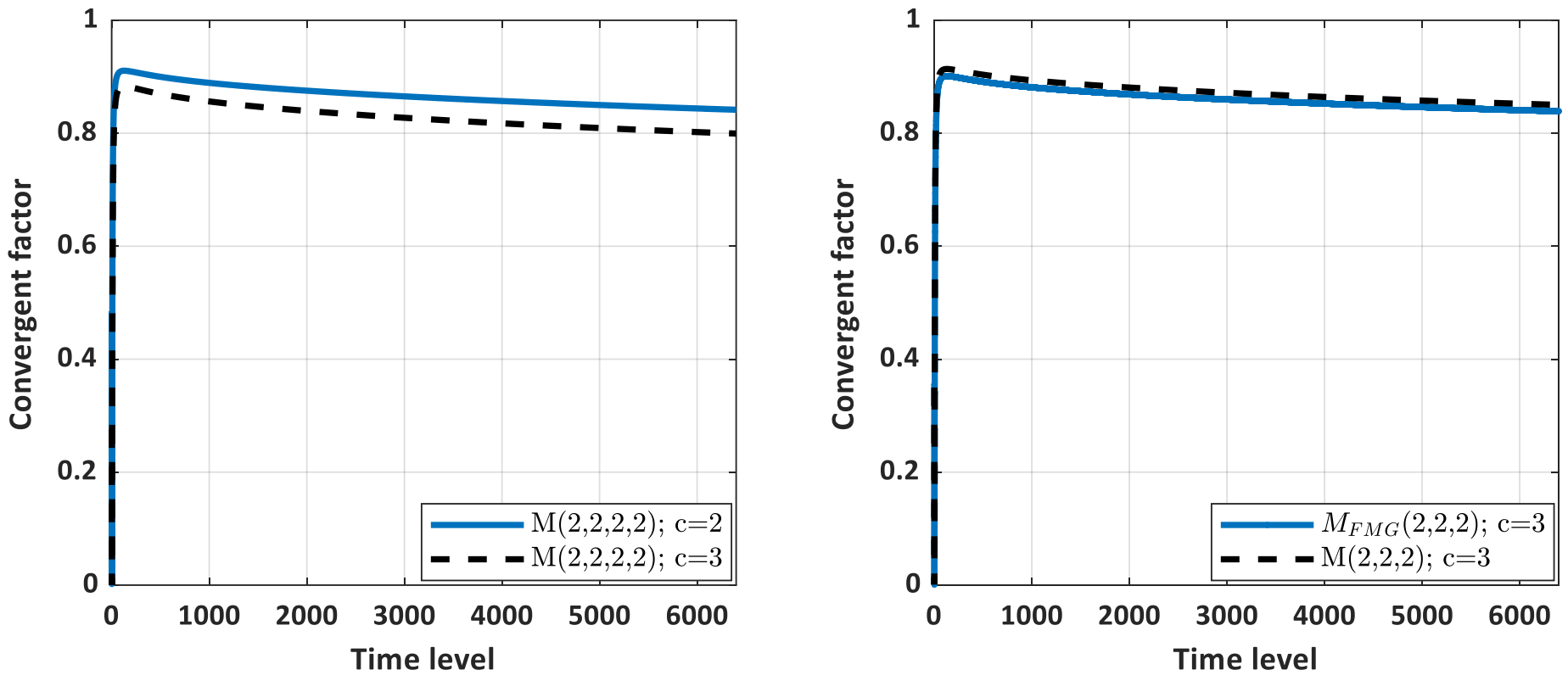

**Fig. 5.** The convergent factor of the multigrid method ($h = 0.0125$).

**Table 3.** Comparison of the global maximum iterations and average CPU time(s) ($h = 0.0125$, outer smoothing = 2 and $c = 3$).

| Method | $M_{FMG}(2,2,2)$ | $M(2,2,2)$ | $M(2,2,2,2)$ | $M(2,2,2,2,2,2)$ | Previous Method |
|---|---|---|---|---|---|
| CPU time(s) | 0.617 | 0.955 | 0.846 | 0.805 | 0.667 |
| Global max. iter. | 94 | 105 | 76 | 60 | 1110 |

**Table 4.** Measured convergent factor when $\tau = T$ with different grid sizes and fixed time step $k = 0.0001$.

| $h$ | 3/80 | 3/96 | 3/120 | 3/160 | 3/192 | 3/240 |
|---|---|---|---|---|---|---|
| $M(2,2,2,2)$ | 0.2823 | 0.3685 | 0.6107 | 0.6107 | 0.6882 | 0.7703 |

To check the convergent rate of our present method, we used the numerical solution of the asset option in regime 1, $T = 0.1$, and three sequence M-cycle with full multigrid initialization. We then defined the maximum error using the notation:

$$E(h,k) = \max_{0\le i\le M} |(u_1)_i^n(h,k) - (u_1)_i^n(h/2,k/4)|, \tag{47a}$$

$$E(h/2,k/4) = \max_{0\le i\le M} |(u_1)_i^n(h/2,k/4) - (u_1)_i^n(h/4,k/16)|, \tag{47b}$$

where $k = h^2$. $(u_1)_i^n(h,k)$, $(u_1)_i^n(h/2,k/4)$, and $(u_1)_i^n(h/4,k/16)$ are the numerical solutions from regime 1 obtained based on $h$, $k/4$, $h/4$, and $k/16$, respectively. As such, the convergent rate was evaluated using the following equation:

$$RoC = \log_2 \frac{E(h,k)}{E(h/2,k/4)}. \tag{48}$$

Results of the maximum error and rate of convergence (RoC) were displayed in Table 5. The rate of convergence presented in Table 5 is around 3.0. The piecewise continuous property of the asset option and the discontinuity of its first derivative at the payoff could contribute to the reduction in the convergent rate. Moreover, it may be because we used the third-order interpolation method for grid correction and interpolation of function in the full multigrid initialization and third-order Hermite function to approximate the coupled delta option in each regime which might further reduce the convergent rate of our numerical scheme. We expect the convergent rate to further increase if we incorporate suitable adaptive or local mesh refinement with the present multigrid strategy for which we will further investigate.

**Table 5.** The maximum errors and convergence rates in space in Regime 1.

| $h$ | maximum error | RoC |
|---|---|---|
| 0.05 | | |
| 0.025 | $2.1 \times 10^{-2}$ | |
| 0.0125 | $2.9 \times 10^{-3}$ | 2.90 |
| 0.00625 | $3.0 \times 10^{-4}$ | 3.20 |

**Example 2:** In this example, to check the effect of varying volatility and interest rate values on the convergence of our numerical solution using the present multigrid algorithm, we use the same data from example 1 except that we consider a smaller volatility and interest rate values for each regime as follows:

$$K = 9, \quad T = 1, \quad Q = \begin{bmatrix} -6 & 6 \\ 9 & -9 \end{bmatrix}, \quad \boldsymbol{r} = \begin{bmatrix} 0.05 \\ 0.05 \end{bmatrix}, \quad \boldsymbol{\sigma} = \begin{bmatrix} 0.15 \\ 0.20 \end{bmatrix}, \quad \varepsilon = 10^{-8}. \tag{49}$$

The results were listed and displayed in Table 6 and Fig. 6, respectively. We observed that our present multigrid method is not sensitive to small volatility and interest values. Moreover, from Fig. 6, we did not observe any oscillation in our numerical solutions especially, in the Greek parameters.

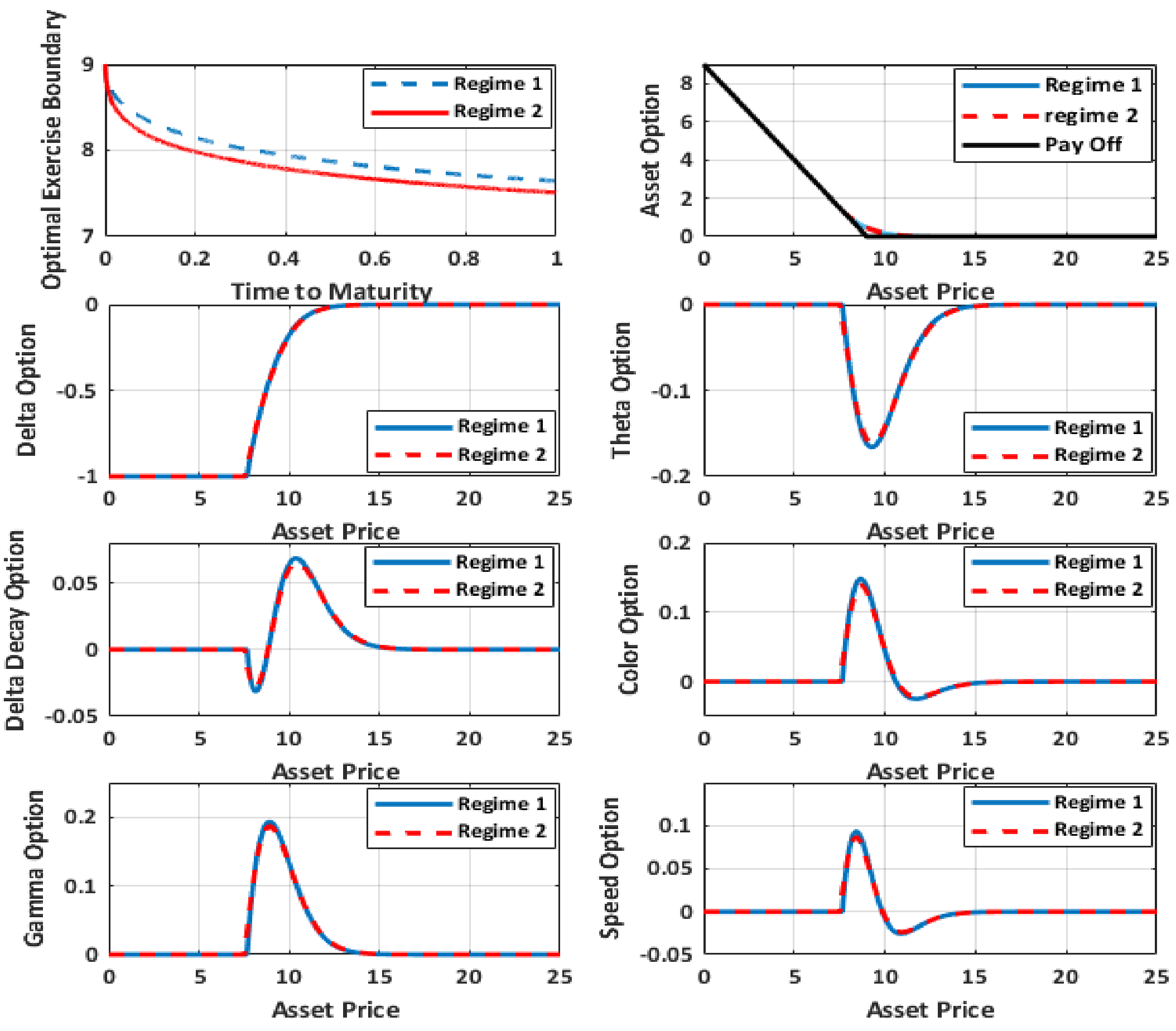

**Fig. 6.** Optimal exercise boundaries, asset options, and option Greeks for example 2 $(\tau = T)$.

**Table 6.** Comparison of the American put option price in regime 1 $(h = 0.0125)$

| S | $M(2,2,2)$ | |
|---|---|---|
| | Regime 1 | Regime 2 |
| 6.0 | 3.0000 | 3.0000 |
| 9.0 | 0.4667 | 0.4615 |
| 12.0 | 0.0165 | 0.0187 |

## 5.2. Four-Regime Example

Consider a four regime-switching example with the strike price $K = 9$ and the expiration time $T = 1$. From previous works of literature, it is challenging to approximate option Greeks accurately beyond two-regime examples. In this example, we focus only on the performance of the M-cycle multigrid method in approximating the option Greeks accurately. We chose the interval $0 \leq x_m \leq 3$ and the time step $k$ was determined using $k = h^2$ with $h = 0.0125$. The remaining parameters were given as follows:

$$Q = \begin{bmatrix} -1 & 1/3 & 1/3 & 1/3 \\ 1/3 & -1 & 1/3 & 1/3 \\ 1/3 & 1/3 & -1 & 1/3 \\ 1/3 & 1/3 & 1/3 & -1 \end{bmatrix}, \qquad \boldsymbol{r} = \begin{bmatrix} 0.02 \\ 0.10 \\ 0.06 \\ 0.15 \end{bmatrix}, \qquad \boldsymbol{\sigma} = \begin{bmatrix} 0.90 \\ 0.50 \\ 0.70 \\ 0.20 \end{bmatrix}. \tag{50}$$

The solution profiles for the asset option, option Greeks, and the optimal exercise boundary for each regime were displayed in Fig. 7.

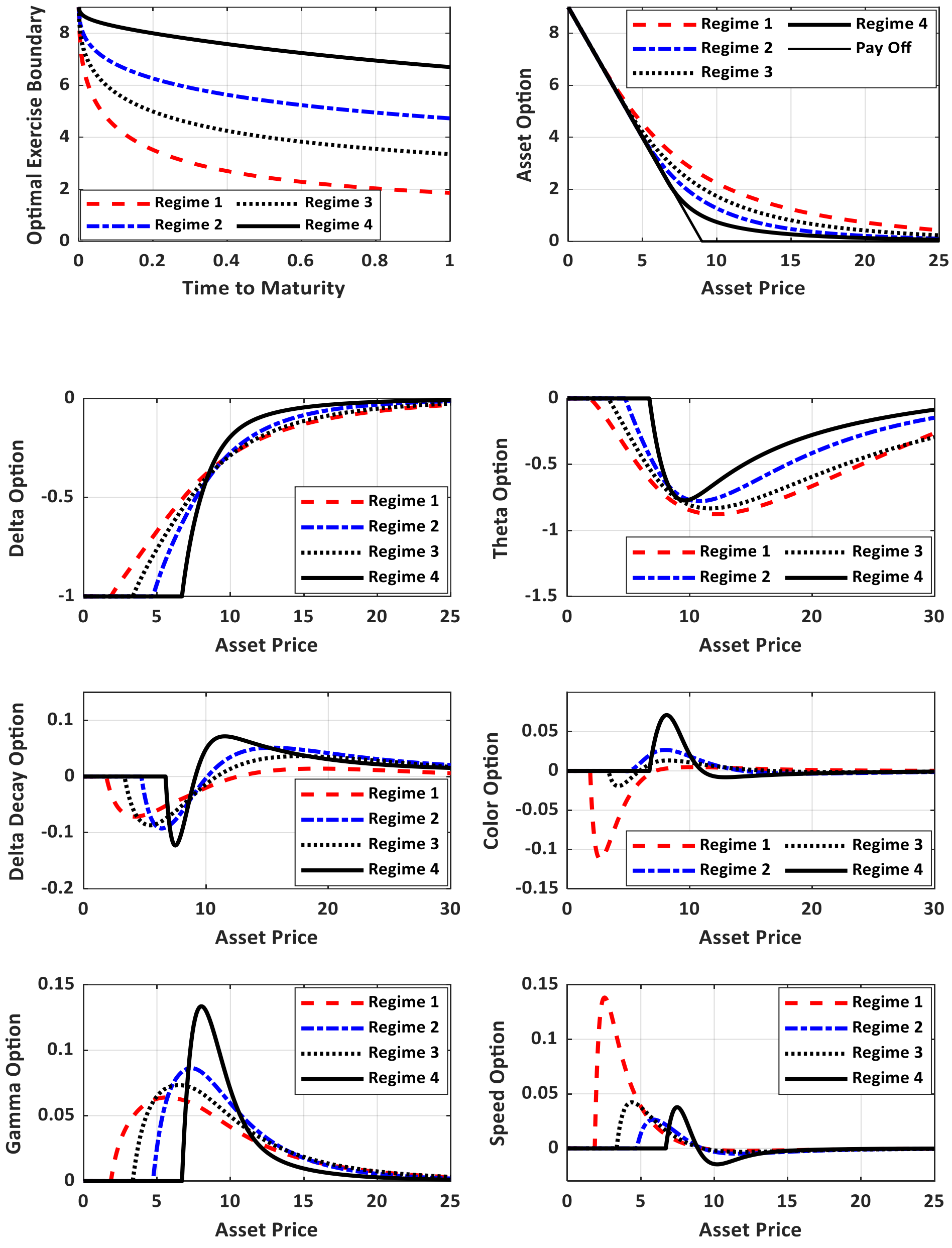

**Fig. 7.** Optimal exercise boundaries, asset options, and option Greeks for the four-regime example using M-cycle multigrid ($\tau = T, c = 3$).

**Table 7.** Comparison of American put options price for the four-regime example ($c = 2, h = 0.0125$).

| | MTree | | | | RBF-FD | | | | $M(2,2,2,2)$ | | | |
|---|---|---|---|---|---|---|---|---|---|---|---|---|
| S | Reg 1 | Reg 2 | Reg 3 | Reg 4 | Reg 1 | Reg 2 | Reg 3 | Reg 4 | Reg 1 | Reg 2 | Reg 3 | Reg 4 |
| 7.5 | 3.1433 | 2.2319 | 2.6746 | 1.6574 | 3.1424 | 2.2320 | 2.6744 | 1.6576 | 3.1417 | 2.2319 | 2.6746 | 1.6576 |
| 9.0 | 2.5576 | 1.5834 | 2.0568 | 0.9855 | 2.5564 | 1.5835 | 2.0566 | 0.9857 | 2.5547 | 1.5835 | 2.0567 | 0.9859 |
| 10.5 | 2.1064 | 1.1417 | 1.6014 | 0.6533 | 2.1052 | 1.1415 | 1.6013 | 0.6554 | 2.1014 | 1.1414 | 1.6012 | 0.6553 |
| 12.0 | 1.7545 | 0.8377 | 1.2625 | 0.4708 | 1.7527 | 0.8377 | 1.2625 | 0.4708 | 1.7466 | 0.8375 | 1.2620 | 0.4706 |

From Fig. 7, the smoothness in the solution of the option Greeks (gamma and speed options) can easily be observed. This is an indication that the present method provides more accurate solutions. Furthermore, we compared our results with the existing methods as listed in Table 7. From Table 7, the results obtained from our M-cycle multigrid algorithm are very close to those obtained from the MTree [29] and RBF-FD [27] methods. Moreover, the plot profile of the gamma option for our method does not exhibit oscillation like the one seen in MOL and RBF-FD methods.

## 6. Conclusion

We have presented a multigrid algorithm for solving American put options with regime-switching and compared its numerical performance with our previous method (that is based on the classical Gauss-Seidel iterative method) and other existing methods. The numerical investigation is carried out with two- and four-regime examples. In the two-regime example, the global maximum iteration required to achieve numerical convergence with $\varepsilon = 10^{-8}$ is much smaller with the multigrid method when compared with our previous method that is based on the Gauss-Seidel iteration. In terms of error per iteration, the error reduces faster in the multigrid method when compared with our previous method. Based on the three grid sequences as presented in Table 3, the M-cycle with full multigrid initialization has less CPU time when compared with the conventional M-cycle multigrid method. In general, the results from the two- and four-regime examples show that the multigrid algorithm presents an efficient tool for pricing American options with regime-switching, especially when it involves a large state space and a more accurate numerical solution is required.